\documentclass[twoside,english,aip,jap,reprint]{revtex4-1}
\usepackage[T1]{fontenc}
\usepackage[latin9]{inputenc}
\usepackage{graphicx}
\usepackage{amssymb}
\usepackage{siunitx}



\usepackage{babel}
\usepackage{url}

\begin{document}

\graphicspath{ {./figure/},{./figure/AllCVD/},{./figure/ExfolhBN/} }

\title{Gate-tunable Hall sensors on large area CVD graphene protected by h-BN with 1D edge contacts}

 \author{Bogdan Karpiak}
 \affiliation{Department of Microtechnology and Nanoscience, Chalmers University
 of Technology, SE-41296, G\"{o}teborg, Sweden}

 \author{Andr{\'e} Dankert}
 \email{andre.dankert@chalmers.se}
 \affiliation{Department of Microtechnology and Nanoscience, Chalmers University
 of Technology, SE-41296, G\"{o}teborg, Sweden}

 \author{Saroj P. Dash}
 \email{saroj.dash@chalmers.se}
 \affiliation{Department of Microtechnology and Nanoscience, Chalmers University
 of Technology, SE-41296, G\"{o}teborg, Sweden}

\begin{abstract}
Graphene is an excellent material for Hall sensors due to its atomically thin structure, high carrier mobility and low carrier density. However, graphene devices need to be protected from the environment for reliable and durable performance in different environmental conditions. Here we present magnetic Hall sensors fabricated on large area commercially available CVD graphene protected by exfoliated hexagonal boron nitride (h-BN). To connect the graphene active regions of Hall samples to the outputs the 1D edge contacts were utilized which show reliable and stable electrical properties. The operation of the Hall sensors shows the current-related sensitivity up to \SI{345}{V/(AT)}. By changing the carrier concentration and type in graphene by the application of gate voltage we are able to tune the Hall sensitivity.
\end{abstract}
\maketitle

\section{Introduction}

As our society becomes more integrated with information technology, sensors are getting increasingly important as can be seen from the consistent global market growth~\cite{marketReportZero}. A variety of different sensors based on magnetic field sensing are used~\cite{MageticSensorsGeneralOverview} with more than 50~$\%$ of which~\cite{marketReportTwo} exploit the Hall effect for their operation. These are utilized in many fields such as healthcare, automotive, industry, and consumer electronics for a broad spectrum of applications including position sensing, current monitoring, proximity detection and others~\cite{MageticSensorsGeneralOverview,GeneralMicroHallDevices,PopovicHallEffectDevices}. Si is the most commonly used material in active regions of Hall sensors as the technology is highly mature with relatively low manufacturing costs and reasonable current-related sensitivities $S_I\sim\SI{100}{V/(AT)}$~\cite{PopovicHallEffectDevices,GeneralMicroHallDevices,VervaekeSizeDep,KejikIntegratedMicroHall}. For  better sensitivity, Hall sensors based on high mobility compound semiconductor heterostructures are used, but yield an increased production price~\cite{KunetsThreeFiveCompoundsBasedHallSensors,HaraGaAsHall,Bando2DelectronGasHallSensors,HeremansOverviewHallSensorsApplications}.

Fostered by the constant strive for performance improvements and market price reduction, new materials are being considered for utilization in Hall sensors. Graphene is a highly promising material for Hall sensing applications due to its beneficial properties such as low charge carrier concentration and high mobility. Recent experimental reports~\cite{XuBatchFabricated,DauberUltraSensitive,HuangUltraSensitive,WangEncapsulatedHall,JooGrapheneOnBN} have already proven the feasibility of graphene Hall sensors with current-related sensitivity more than 60 times higher than in silicon-based sensors~\cite{DauberUltraSensitive}. Furthermore, graphene's 2D nature allows its application in flexible and transparent electronics.

However, the high-performance proof-of-concept devices demonstrated so far either use exfoliated graphene flakes~\cite{DauberUltraSensitive}, are made of a manually selected single crystal CVD graphene area~\cite{WangEncapsulatedHall}, or the graphene active layer is not protected from external influences required for robust device operation~\cite{XuBatchFabricated,JooGrapheneOnBN,HuangUltraSensitive,WangFlexibleHall}. To harness the advantages of industrially compatible large area CVD graphene in Hall sensors, it is necessary to protect the graphene layer for reliable device performance in different environmental conditions. 

Here we report magnetic Hall elements fabricated on large area CVD graphene covered by hexagonal boron nitride (h-BN) for protection against the environmental influences to ensure the reliable operation and long lifetime. The insulating and two-dimensional nature of the h-BN capping layer is expected to have a good interface with graphene, containing fewer dangling bonds and charge traps~\cite{DeanBoronNitrideSubstrates}.
Our devices incorporate 1D edge contacts to graphene/h-BN heterostructures, which circumvent the effects related to contact-induced doping. These factors are crucial for reliable performance of graphene Hall sensors in an ambient environment.

\section{Results and discussions}

\begin{figure}[!b]
\centering
\includegraphics[trim=0mm 0mm 0mm 0mm, clip=true,width=\linewidth]{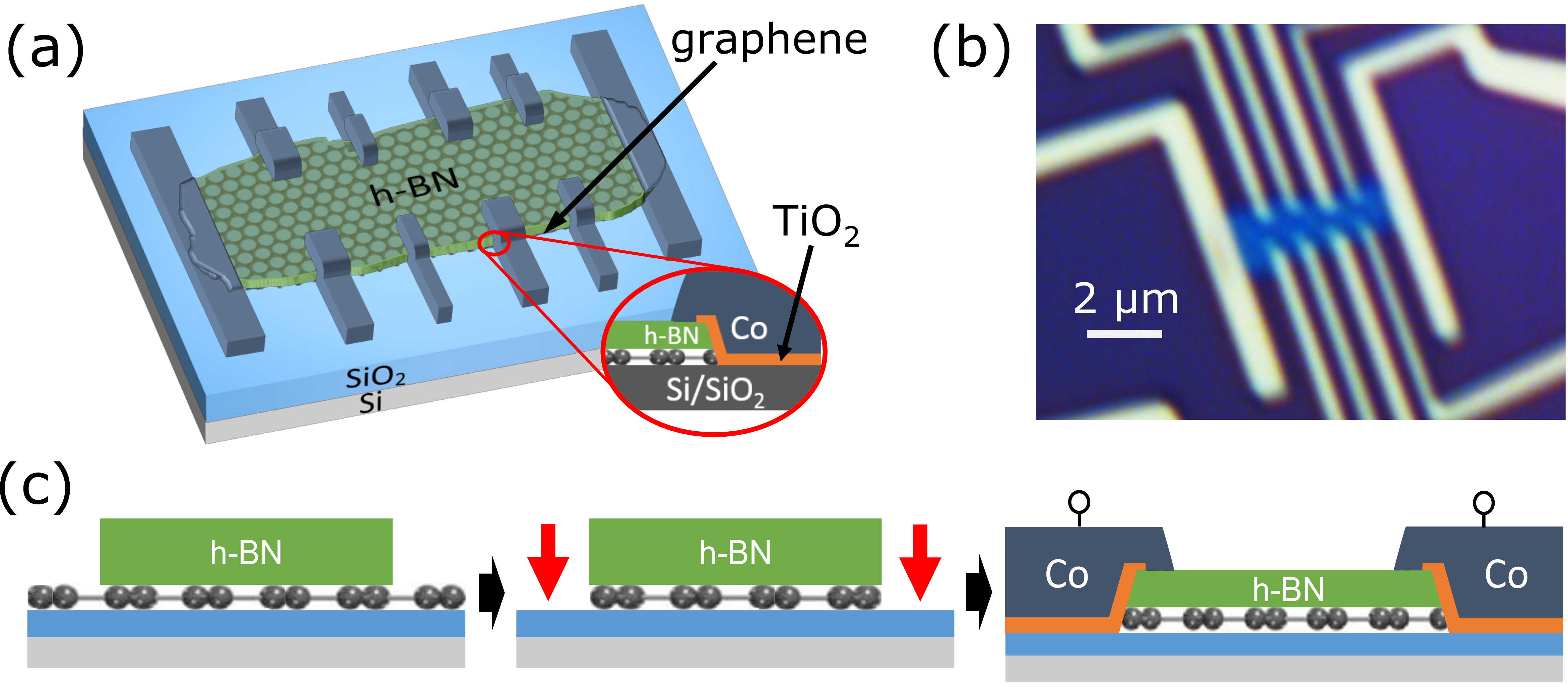}
\vspace{-0.3cm}
\caption{Hall sensor fabrication with CVD graphene/h-BN heterostructures with 1D edge contacts. (a)~Schematic representation and (b)~ optical microscope picture of the fabricated device. (c)~Schematics of the fabrication process steps from left to right: the preparation of graphene/h-BN heterostructures; patterning by oxygen plasma; and deposition of 1D edge contacts. Dimensions in (a) and (c) are not to the scale.  \label{fig:fig1}}
\end{figure}

The Hall sensor devices were fabricated on commercially available CVD graphene after it was transferred on \mbox{Si/SiO$_2$} substrate (Graphenea~\cite{grapheneaCompany}). The schematic illustration and an optical image of the investigated Hall samples are shown in Fig.~\ref{fig:fig1}a and Fig.~\ref{fig:fig1}b respectively. The micro-fabrication process steps are depicted in Fig.~\ref{fig:fig1}c. The h-BN was exfoliated from bulk crystal on top of graphene by means of a regular scotch-tape method. Then the unprotected graphene regions were etched away with oxygen plasma.\@ 1D edge contacts~\cite{WangOneDcontact} were fabricated by means of electron beam lithography and electron beam evaporation of metals (\SI{1}{\nano\metre}~TiO$_2$/\SI{65}{\nano\metre}~Co) followed by liftoff in acetone/isopropyl alcohol.

\begin{figure}[!b]
\centering
\includegraphics[trim=0mm 0mm 0mm 0mm, clip=true,width=0.8\linewidth]{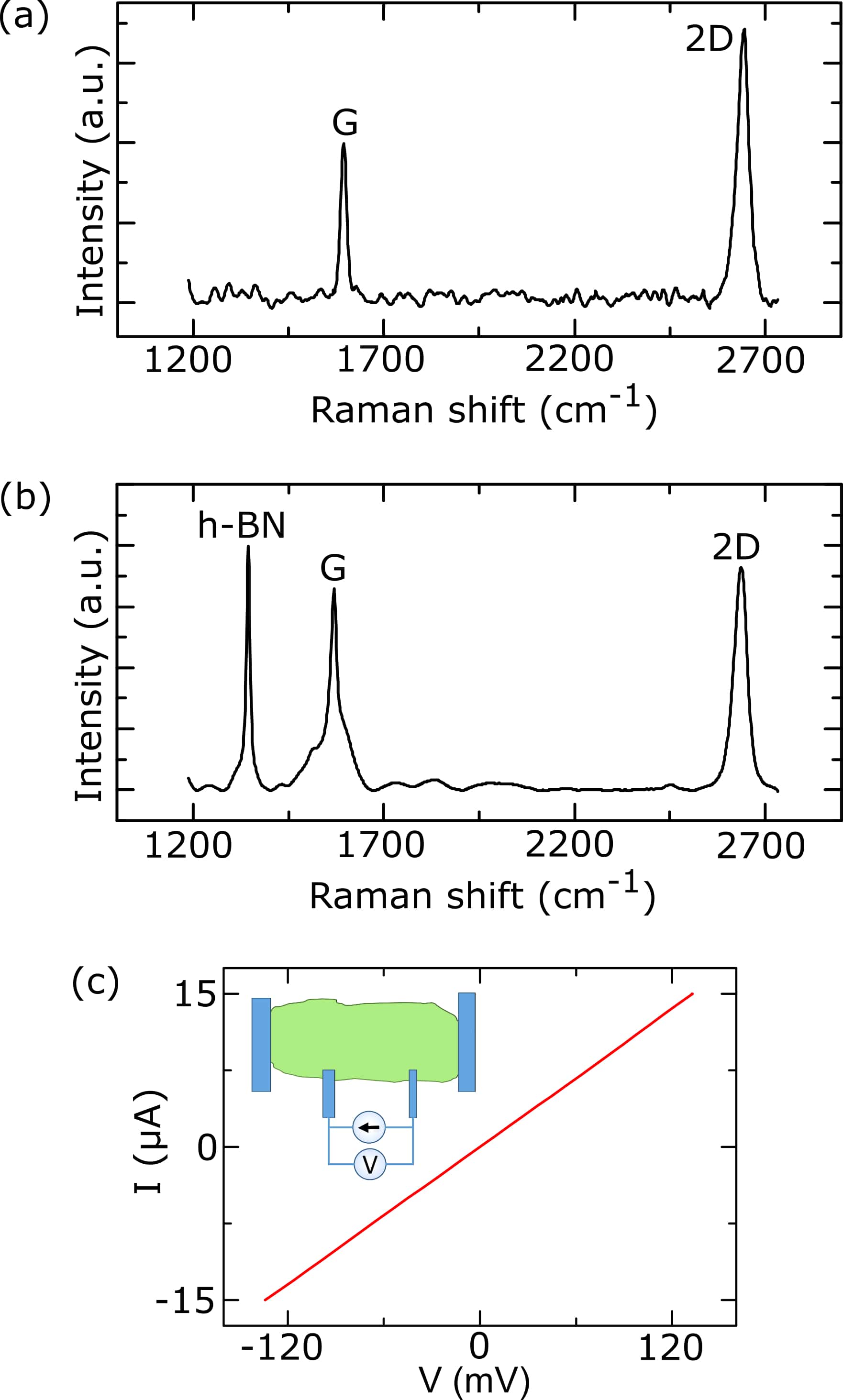}
\caption{Characterization of CVD graphene/h-BN heterostructures and 1D edge contacts. (a)~Raman spectrum of CVD graphene and (b) graphene/h-BN heterostructure. (c)~Typical two-terminal I-V characteristic of the 1D edge contacts to graphene at room temperature. Inset: two-terminal measurement configuration. \label{fig:fig2}}
\end{figure}

Fig.~\ref{fig:fig2} shows the characterization of graphene/h-BN heterostructure and 1D edge contacts to graphene. The Raman spectra of CVD graphene and the graphene/h-BN heterostructure~\cite{FerrariRamanGrapheneLayers,HuntingBN} are shown in Fig.~\ref{fig:fig2}a and Fig.~\ref{fig:fig2}b respectively. The absence of a band splitting of the 2D peak and its higher intensity compared to the G peak indicate that the graphene in the heterostructure is a monolayer~\cite{FerrariRamanGrapheneLayers}. The CVD graphene used here has grain sizes of 1-3 $\mu$m (Graphenea). The typical two-terminal current-voltage (I-V) characteristic at room temperature (Fig.~\ref{fig:fig2}c) shows a linear behaviour and the contact resistances, estimated from the analysis of data from local four- and two-terminal measurements, are reproducible in several devices. The field-effect mobility of the measured graphene at room temperature is found to be $\sim\SI{133}{cm^2/V.s^{-1}}$ and sheet resistance $\sim\SI{17}{\kilo\ohm/}\square$.

\begin{figure}[!t]
\centering
\includegraphics[trim=0mm 0mm 0mm 0mm, clip=true,width=0.9\linewidth]{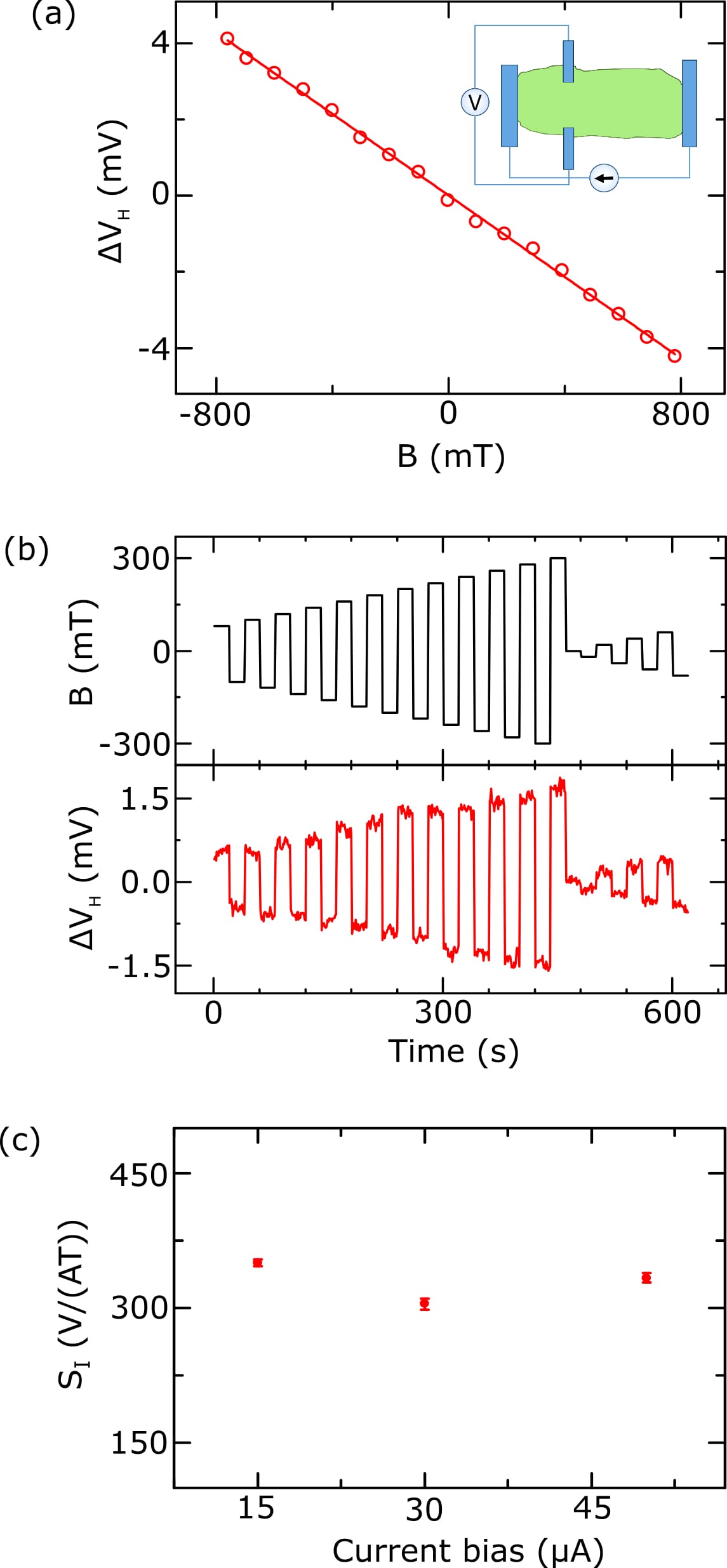}
\caption{Operation of CVD graphene/h-BN Hall elements. (a)~Output Hall voltage as a function of perpendicular magnetic field measured at $I=\SI{15}{\micro\ampere}$ at room temperature (circles) with linear fitting (solid line) according to Eq.~\ref{eq:HallMain}. Inset: Hall measurement configuration. (b)~Output Hall voltage as a function of time at different applied magnetic fields measured at $I=\SI{15}{\micro\ampere}$ and at room temperature. (c)~Current-related sensitivity as a function of current bias. \label{fig:fig3}}
\end{figure}

\begin{figure*}[!t] 
\centering
\includegraphics[trim=0mm 0mm 0mm 0mm, clip=true,width=\linewidth]{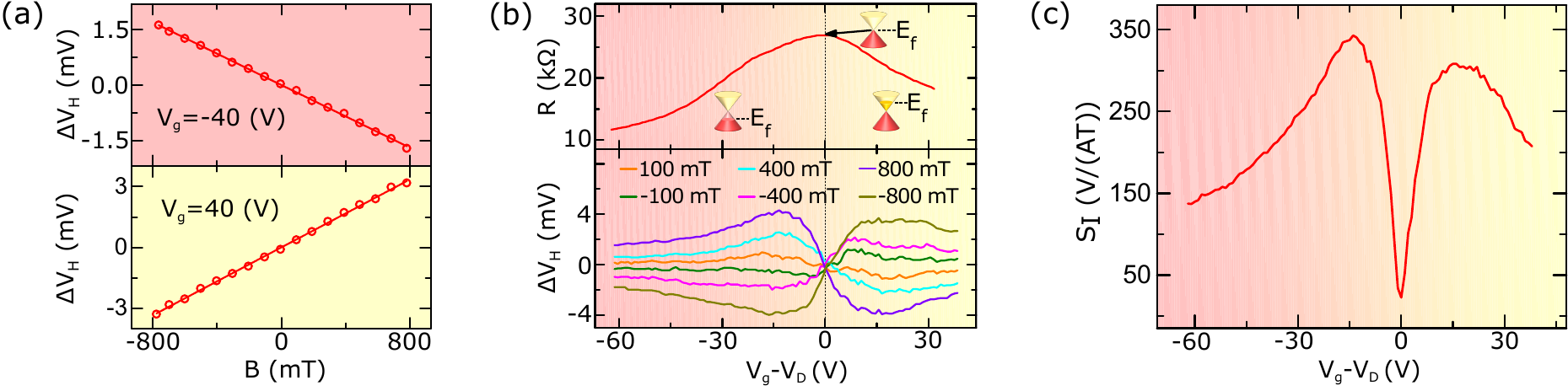}
\caption{Gate tunability of CVD graphene/h-BN Hall sensors. (a)~Output Hall voltage as a function of perpendicular magnetic field at gate voltages of $\pm \SI{40}{V}$. (b)~Back gate dependence of graphene resistance (top) and Hall voltage response at different perpendicular magnetic fields (bottom). V$_D$ is the gate voltage corresponding to the Dirac point. (c)~Absolute value of current-related sensitivity calculated from Hall voltage response as a function of gate voltage according to Eq.~\ref{eq:SI}. \label{fig:fig4}}
\end{figure*}

Fig.~\ref{fig:fig3}a shows the Hall measurements performed using the measurement configuration depicted in the inset. The Hall voltage ($V_H$) response of investigated samples obtained during the magnetic field (B) sweep at applied current bias $I=\SI{15}{\micro\ampere}$ is fitted with~\cite{HeremansOverviewHallSensorsApplications,PopovicHallEffectDevices} 
\begin{equation}
    \frac{\partial V_H}{\partial B}=\frac{1}{en_{2D}}I, \label{eq:HallMain}
\end{equation}
where $e$ is the elementary electron charge and $n_{2D}$ is the charge carrier density. Graphene was found to exhibit hole conduction at zero back gate voltage with a sheet charge carrier concentration of \mbox{$n_{2D}=\SI{1.75d12}{cm^{-2}}$} and background voltage offset of \SI{3}{\milli\volt}, which has been subtracted from the measured raw data. The linearity errors~\cite{PopovicHallEffectDevices,HaraGaAsHall,XuBatchFabricated} were found to be within $\pm 3.1~\%$ with an average absolute value of \mbox{1.3~\%} over a large magnetic field range from \SIrange{-760}{780}{\milli\tesla} at room temperature. From the measured Hall voltage response as a function of time at different perpendicular magnetic fields (Fig.~\ref{fig:fig3}b) one can estimate the noise level and the minimum resolvable magnetic field of \mbox{\SI{\sim 20}{\milli\tesla}} at room temperature. From the Hall measurements, the calculated current-related Hall sensitivity
\begin{equation}
    S_I=\left. \frac{1}{I}\frac{\partial V_{H}}{\partial B} \right|_{I=const} \label{eq:SI}
\end{equation}
did not show significant bias-related change in the bias current range from \SIrange{15}{50}{\micro\ampere} at room temperature~(Fig.~\ref{fig:fig3}c).

Next, we investigated the Hall sensitivity of graphene for different carrier concentrations in electron- and hole-doped regimes (Fig.~\ref{fig:fig4}). Graphene has the unique property that its charge carrier type and concentration can be tuned continuously by applying gate voltage. Fig.~\ref{fig:fig4}a shows the Hall measurements with application of gate voltages \mbox{$V_{g}=\pm\SI{40}{V}$}, where a  sign change is observed in the slope of Hall response for electron- and hole-doped regimes. From the back gate dependence of the graphene resistance the charge neutrality point was found at \mbox{$V_{g}=\SI{26}{V}$}. Such full gate-dependent Hall effect measurements were performed by sweeping the back gate voltages from electron to hole type of conduction across the Dirac point in the presence of different perpendicular magnetic fields (bottom panel of Fig.~\ref{fig:fig4}b). To reduce the influence of the device geometry, the Hall response at different magnetic fields ($V_H(B)$) is subtracted from measured response at \SI{0}{\tesla} magnetic field ($V_{H0}$): $\Delta V_H=V_H(B)-V_{H0} \label{eq:VhSubtractReferenceZeroTesla}$~\cite{DauberUltraSensitive}. Here, we observe a change in the amplitude and sign of $\Delta V_H$ by sweeping the gate voltage. Using Eq.~\ref{eq:SI} we extracted the current-related sensitivity by fitting the \mbox{$\Delta V_H$-$B$} dependencies. It is plotted in Fig.~\ref{fig:fig4}c as a function of gate voltage at room temperature. This dependence reveals the gate tuning of the sensitivity with maxima (up to \SI{345}{V/(AT)}) close to the graphene Dirac point. Such tunability of Hall response stems from gate voltage-induced change of the carrier concentration in the graphene sheet yielding a change of the current-related sensitivity and the output Hall voltage (Eq.~\ref{eq:HallMain} and Eq.~\ref{eq:SI}).

These sensitivities of the large area CVD graphene/h-BN heterostructure-based Hall sensors are at least three times higher than current silicon-based devices~\cite{PopovicHallEffectDevices,GeneralMicroHallDevices,VervaekeSizeDep,KejikIntegratedMicroHall}. Previous studies reported on the performance of unencapsulated graphene Hall sensors~\cite{WangFlexibleHall,XuBatchFabricated,HuangUltraSensitive} with sensitivities up to 2093 V/(AT). Such unprotected devices are known to be extremely sensitive to environmental parameters and degrade rapidly. Consequently, an insulating barrier is required to protect the graphene layer. Even though recent studies on Al$_2$O$_3$ show very promising results on easily scalable and reproducible device fabrication techniques~\cite{AloxideEncapsulation}, the close nature of insulating h-BN demonstrated outstanding electronic properties when encapsulating graphene~\cite{DeanBoronNitrideSubstrates}. Such exfoliated h-BN/graphene/h-BN heterostructure Hall devices were demonstrated with sensitivities 15 times higher than in our h-BN capped CVD graphene~\cite{DauberUltraSensitive}. This can be attributed to growth quality of CVD graphene, presence of grain boundaries and charge doping from SiO$_2$ substrate as well as contaminations introduced during wet transfer of CVD graphene. Contaminations and grain boundaries are also likely reasons behind the observed low mobility and high sheet resistance of graphene. The measured Hall sensitivities could be further increased by improving the CVD graphene mobility by reducing the SiO$_2$ substrate-induced doping and graphene-substrate interactions~\cite{ParkSubstrateModification,XuBatchFabricated} by fully encapsulating CVD graphene in h-BN. Utilizing high-quality, large-grain graphene growth and Ohmic instead of tunnel contacts would also allow to significantly improve graphene characteristics and sensing performance. Furthermore, as a next step forward a fully-scalable fabrication approach should be considered with utilization of only all-CVD \mbox{h-BN/graphene/h-BN} stacks patterned on large area.

\section{Conclusions}

In summary, we demonstrated the operation of graphene magnetic Hall elements fabricated on large area CVD graphene on Si/SiO$_2$ substrate with h-BN capping and 1D edge contacts. Such heterostructure devices showed reliable contact properties and Hall sensor performance. The samples showed a constant bias dependence of the current-related sensitivity and a minimum magnetic field resolution of \mbox{$\sim\SI{20}{\milli\tesla}$} at room temperature. Gate voltage-induced tunability of the Hall response was observed with a maximum current-related sensitivity of \SI{345}{V/(AT)} close to the Dirac point. These results obtained in graphene Hall sensors with h-BN protection are promising for operation in ambient conditions and for their potential application in transparent and flexible electronics.

\section*{Acknowledgements}
We acknowledge financial support from Swedish Research Council, EU Graphene Flagship, EU FlagEra projects, and Graphene center and AoANano program at Chalmers.

\end{document}